\documentclass[]{interact}

\usepackage{amsmath,amssymb,amsfonts}
\usepackage{booktabs}
\usepackage{graphicx}
\usepackage{multirow}
\usepackage{array}
\usepackage{tabularx}
\usepackage{placeins}
\usepackage{url}
\usepackage{hyperref}
\usepackage{algorithm}
\usepackage{algpseudocode}

\usepackage[numbers,sort&compress]{natbib}

\hypersetup{
  colorlinks=true,
  linkcolor=blue,
  citecolor=blue,
  urlcolor=blue,
  pdftitle={Hyperedge-Triggered Hawkes Processes for Diagnosing Higher-Order Temporal Dependence in Event Streams},
  pdfauthor={Zihan Xu}
}

\newcommand{\HTH}{\mathrm{HTH}}
\newcommand{\PW}{\mathrm{PW}}
\newcommand{\E}{\mathcal{E}}
\newcommand{\Hcal}{\mathcal{H}}

\newcommand{\one}{\mathbf{1}}

\newcolumntype{Y}{>{\raggedright\arraybackslash}X}

\newcommand{\safefigure}[2]{%
\IfFileExists{#1}{%
  \includegraphics[#2]{#1}%
}{%
  \fbox{\parbox[c][0.25\textheight][c]{0.85\linewidth}{\centering Missing figure file:\\ \texttt{#1}}}%
}%
}

\begin{document}

\title{Hyperedge-Triggered Hawkes Processes for Diagnosing Higher-Order Temporal Dependence in Event Streams}

\author{\name{Zihan Xu}
\affil{School of Mathematics and Statistics, Qingdao University, Qingdao, China}}

\history{Revised August 2026}
\maketitle

\begin{abstract}
\noindent Monte Carlo simulation is used to assess a likelihood-based EM procedure for diagnosing higher-order temporal dependence in multivariate event streams.\par\medskip
Multivariate Hawkes processes provide a likelihood-based framework for
modelling self- and cross-excitation in continuous-time event streams,
but their standard pairwise formulation can confound dense dyadic
excitation with genuinely group-triggered temporal dependence. We
introduce a hyperedge-triggered Hawkes model in which a candidate group
contributes additional intensity only after all of its member nodes fire
within a prescribed trigger window. A most-recent-anchor convention makes
this higher-order component non-accumulating and leads to a corresponding
piecewise compensator in the likelihood.

The model is estimated by a latent-branching EM algorithm that assigns
event-level responsibilities to background, pairwise, and
hyperedge-triggered sources. Controlled simulations evaluate recovery,
regularisation, convergence, separation from pairwise baselines,
trigger-window and kernel-timescale sensitivity, computational scaling,
and comparison with a parameter-matched third-order surrogate. The
results show that higher-order excitation can be recovered in
informative regimes, while pairwise--hyperedge confounding remains the
main statistical limitation. Applications to retinal and visual-cortex
spike trains are presented as exploratory predictive analyses: the
hyperedge-triggered component can improve candidate-penalised held-out
fit, but the fitted effects should be interpreted as diagnostics of
higher-order temporal dependence rather than calibrated discoveries of
individual biological hyperedges.
\end{abstract}

\begin{keywords}
Hawkes process; EM algorithm; maximum-likelihood estimation;
higher-order temporal dependence; identifiability; neural spike trains
\end{keywords}

\section{Introduction}

Multivariate event streams are often modelled through conditional
intensities that depend on the observed past. Hawkes processes are a
classical and widely used example: past events increase, or otherwise
modulate, the intensity of future events through history-dependent
kernels \citep{Hawkes1971,Ogata1978,Ozaki1979}. In the multivariate
case, the usual specification is dyadic. An event on one component
contributes a decaying term to the intensity of another component. This
pairwise structure is computationally convenient and has made Hawkes
models useful in seismology, finance, social systems, and neural
spike-train analysis.

Dyadic excitation is not always the most natural description of an
event pattern. In neural population data, for example, firing history,
ensemble activity, synchrony, and correlated population states have long
been studied with point-process and population models
\citep{Truccolo2005,Schneidman2006,Pillow2008,Kass2011}. A future event
may be better predicted after several nodes have fired within a short
time interval than after any one of them fires alone. Such a pattern is
temporal rather than purely static: the order and timing of the events
that complete the group matter. At the same time, it is not automatically
evidence of a distinct biological higher-order mechanism. Weak or dense
pairwise effects can also create collective activity patterns
\citep{Schneidman2006}. Separating pairwise excitation from
group-triggered temporal dependence is therefore a statistical modelling
problem.

Higher-order interactions have been studied in hypergraphs, simplicial
complexes, and related network models
\citep{Giusti2016,Battiston2020,Zhang2023HigherOrder}. These approaches
make clear why pairwise representations may be insufficient, but many of
them are static, binned, or not formulated through a continuous-time
point-process likelihood. Conversely, the Hawkes literature provides
well-developed tools for likelihood-based estimation, branching
representations, sparse multivariate modelling, and graphical
interpretation
\citep{VeenSchoenberg2008,LewisMohler2011,Zhou2013,Hansen2015,Bacry2020SparseLowRank,Xu2016,Eichler2017,EmbrechtsKirchner2018}.
The gap addressed in this paper is specific: how can a
group-completion effect be incorporated into a continuous-time Hawkes
likelihood, and when can such an effect be distinguished empirically
from ordinary pairwise excitation?

We study a hyperedge-triggered Hawkes (HTH) process. The model retains
the ordinary pairwise Hawkes component and adds a candidate hyperedge
component. A candidate hyperedge becomes active when all of its member
nodes have fired within a prescribed trigger window. The activation time
is the time at which the pattern is completed. Unlike a linear Hawkes
term, the hyperedge component does not accumulate over all past
completions. Instead, it uses a most-recent-anchor convention: at any
time, only the latest completion of a candidate hyperedge is active, and
a new completion supersedes the earlier one. This convention gives the
higher-order term a short-window pattern-completion interpretation.

The most-recent-anchor convention changes the likelihood calculation.
For an ordinary exponential Hawkes component, the compensator integrates
the contribution of each past event to the end of the observation
window. For the HTH component, this would be wrong: once a later
completion of the same candidate hyperedge occurs, the earlier anchor no
longer contributes. Its exposure must be integrated only until the next
completion, or until the end of the observation window if there is no
later completion. The resulting piecewise compensator is the main
technical ingredient of the likelihood and of the EM updates.

Estimation is based on a latent-branching representation. Each event is
assigned responsibilities for three source classes: background,
ordinary pairwise Hawkes excitation, and hyperedge-triggered excitation.
For independently parameterised candidate hyperedges, the resulting
M-step gives closed-form updates for the baseline rates, pairwise
weights, and hyperedge weights. When the hyperedge weights are
constrained by a non-negative CP parameterisation, the same expected
complete-data objective is maximised by block-coordinate ascent, giving
a generalised EM implementation.

The empirical study is organised around identifiability and predictive
separation. Synthetic simulations are used because the data-generating
mechanism is then known. They evaluate parameter recovery,
regularisation, convergence from random initialisations, separation of
null and signal settings, trigger-window sensitivity,
kernel-timescale effects, computational scaling, and comparison with a
parameter-matched third-order surrogate. The real neural spike-train
analyses are presented more cautiously. They examine whether the HTH
component improves held-out predictive fit after a simple
candidate-count penalty and how sensitive that improvement is to the
candidate set. They are not treated as calibrated discoveries of
individual biological hyperedges. This distinction is important because
formal testing for non-negative Hawkes interactions is already
non-standard in ordinary multivariate settings, and the present setting
also involves data-dependent candidates and a pattern-completion source
class \citep{Lotz2024,BonnetTesting2025}.

The paper makes four contributions.
\begin{enumerate}
  \item We formulate a continuous-time hyperedge-triggered Hawkes model
        in which a short-window completion of a candidate group adds a
        non-accumulating higher-order intensity component.
  \item We derive the piecewise compensator induced by the
        most-recent-anchor convention and incorporate it into the
        likelihood and EM updates.
  \item We develop a latent-branching EM estimator with closed-form
        updates for independently parameterised candidate hyperedge
        weights and a generalised EM extension under a non-negative CP
        constraint.
  \item We use controlled simulations and exploratory neural
        spike-train applications to clarify when higher-order temporal
        dependence is empirically separable from pairwise Hawkes
        excitation and when such interpretation is fragile.
\end{enumerate}

The rest of the paper is organised as follows. Section~\ref{sec:model}
defines the HTH model and its compensator. Section~\ref{sec:em}
derives the latent-branching EM algorithm. Section~\ref{sec:simulation-design}
describes the simulation design and evaluation metrics.
Section~\ref{sec:simulation-results} reports the synthetic experiments.
Section~\ref{sec:real-data} presents the exploratory neural
applications. Section~\ref{sec:discussion} discusses interpretation,
limitations, and practical use.
\section{Hyperedge-triggered Hawkes model}
\label{sec:model}

\subsection{Event notation and conditional intensity}

Consider a marked event sequence observed on the interval $[0,T]$.
Event $i$ occurs at time $t_i$ and on node $n_i\in\{1,\ldots,N\}$, with
$0<t_1<\cdots<t_m\leq T$. Let $\Hcal_t$ denote the event history before
time $t$. For node $n$, the hyperedge-triggered Hawkes process has
conditional intensity
\begin{equation}
\lambda_n(t\mid \Hcal_t)
=
\mu_n
+
\sum_{j:t_j<t}
\alpha_{n_j\to n}\,\phi(t-t_j)
+
\sum_{e\in\E_{\rm cand}:n\in e}
\one\{C_e(t)\neq \varnothing\}
\alpha_e\,\phi\{t-t^\star_e(t)\}.
\label{eq:hth-intensity}
\end{equation}
Here $\mu_n>0$ is the background rate,
$\alpha_{a\to n}\geq0$ is the ordinary pairwise Hawkes weight from
source node $a$ to target node $n$, and $\alpha_e\geq0$ is the weight of
candidate hyperedge $e$. Throughout the paper the kernel is exponential,
\begin{equation}
\phi(u)=\exp(-\beta u)\one\{u>0\},
\label{eq:exp-kernel}
\end{equation}
with a fixed decay rate $\beta>0$. The candidate set $\E_{\rm cand}$ is
a finite collection of node groups. It is treated as part of the fitted
model, not as the output of a calibrated support-recovery procedure.

The first two terms in \eqref{eq:hth-intensity} form an ordinary
multivariate Hawkes component. The third term is different. It is not a
sum over all past completions of a hyperedge. At any time $t$, only the
most recent completion of a candidate hyperedge is active.

\subsection{Pattern completion and most-recent anchor}

Let $e\subseteq\{1,\ldots,N\}$ be a candidate hyperedge and let
$\Delta>0$ be the trigger window. A completion of $e$ is defined at an
observed event time. Specifically,
\begin{equation}
C_e
=
\left\{
t_i:
n_i\in e,\;
\text{for every } v\in e \text{ there exists } k\leq i
\text{ with } n_k=v
\text{ and } t_k\in[t_i-\Delta,t_i]
\right\}.
\label{eq:completion-set}
\end{equation}
Thus $t_i\in C_e$ when the event at $t_i$ completes the presence of all
members of $e$ within the window $[t_i-\Delta,t_i]$. For $t>0$, define
\begin{equation}
C_e(t)=\{c\in C_e:c<t\},
\qquad
t^\star_e(t)=\max C_e(t)
\label{eq:anchor}
\end{equation}
whenever $C_e(t)$ is non-empty. If no completion of $e$ has occurred
before $t$, the indicator in \eqref{eq:hth-intensity} sets the
hyperedge contribution to zero.

This construction gives the hyperedge term a pattern-completion
interpretation. A group-level excitation begins when the pattern is
completed and then decays with the exponential kernel. If the same
pattern is completed again, the new completion becomes the active
anchor. Earlier anchors are therefore superseded. This convention
distinguishes the present hyperedge term from an accumulated linear
Hawkes term.

\begin{figure}[tbp]
\centering
\safefigure{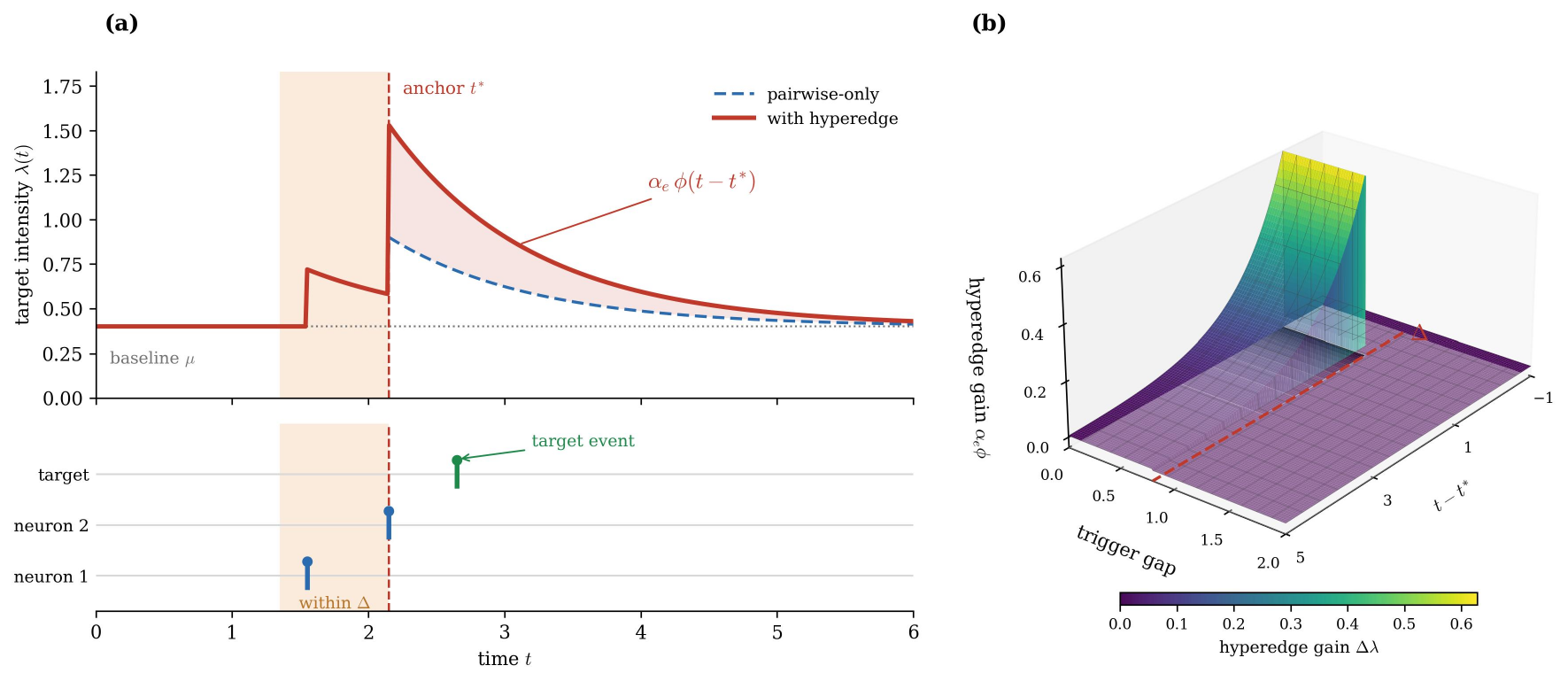}{width=0.95\textwidth}
\caption{Hyperedge-triggered activation. A candidate hyperedge becomes
active when its member nodes complete a firing pattern within the
trigger window. Under the most-recent-anchor convention, a later
completion supersedes the earlier anchor, so the hyperedge contribution
is a single decaying term rather than an accumulating sum over all past
completions.}
\label{fig:hth-mechanism}
\end{figure}

\subsection{Observed-data log-likelihood}

For parameters
$\theta=(\mu,\alpha^{\rm pw},\alpha^{\rm he})$, the observed-data
log-likelihood on $[0,T]$ is
\begin{equation}
\ell(\theta)
=
\sum_{i=1}^{m}\log \lambda_{n_i}(t_i\mid \Hcal_{t_i})
-
\sum_{n=1}^{N}\int_0^T \lambda_n(t\mid \Hcal_t)\,dt .
\label{eq:observed-loglik}
\end{equation}
The baseline contribution to the compensator is $\sum_n \mu_n T$. The
ordinary pairwise Hawkes contribution is
\begin{equation}
\sum_{n=1}^{N}
\sum_{j=1}^{m}
\alpha_{n_j\to n}
\frac{1-\exp\{-\beta(T-t_j)\}}{\beta}.
\label{eq:pairwise-compensator}
\end{equation}
The hyperedge contribution requires separate treatment because of the
most-recent-anchor convention.

\subsection{Piecewise hyperedge compensator}

Let the completion times of candidate hyperedge $e$ be
\[
c_1<c_2<\cdots<c_{M_e},
\]
and set $c_{M_e+1}=T$. Between $c_r$ and $c_{r+1}$, the active anchor of
$e$ is $c_r$. The exposure of a single target node in $e$ is therefore
\begin{equation}
\Lambda_e
=
\sum_{r=1}^{M_e}
\int_{c_r}^{c_{r+1}}
\exp\{-\beta(t-c_r)\}\,dt
=
\sum_{r=1}^{M_e}
\frac{1-\exp\{-\beta(c_{r+1}-c_r)\}}{\beta}.
\label{eq:hyperedge-exposure}
\end{equation}
Since the same candidate hyperedge contributes to the intensity of each
member node, its total exposure in the likelihood is $|e|\Lambda_e$.

The point of \eqref{eq:hyperedge-exposure} is not merely computational.
A naive compensator that integrates the kernel from every completion
time to $T$ would keep old completions active after they have been
superseded by later completions. That would overcount exposure and
distort the hyperedge weight update. Equation
\eqref{eq:hyperedge-exposure} is the likelihood counterpart of the
most-recent-anchor definition in \eqref{eq:anchor}. Exact compensator
calculations are central in likelihood-based Hawkes estimation, and here
the relevant compensator is determined by the non-accumulating anchor
mechanism rather than by the ordinary accumulated Hawkes structure
\citep{Ozaki1979,Bonnet2021}.

With this notation, the hyperedge part of the integrated intensity is
\begin{equation}
\sum_{e\in\E_{\rm cand}} \alpha_e |e| \Lambda_e .
\label{eq:hyperedge-compensator-total}
\end{equation}

\subsection{Non-negative CP parameterisation}

When candidate hyperedges are parameterised independently, each
$\alpha_e$ is a separate non-negative parameter. This representation is
useful for deriving the EM updates and for sparse candidate sets, but it
does not scale well if many higher-order groups are considered. For
$K$-node hyperedges, we also use a non-negative CP parameterisation
\citep{KoldaBader2009},
\begin{equation}
\alpha_e
=
\sum_{r=1}^{R}
\prod_{v\in e} F_{vr},
\qquad
F_{vr}\geq 0 ,
\label{eq:cp-param}
\end{equation}
where $R$ is a chosen rank. For fixed $R$, this reduces the number of
free hyperedge parameters from combinatorial order in $N$ to order
$NR$.

The CP parameterisation is used here as a structured dimension-reduction
device. It should not be interpreted as a guarantee that the factors
recover biological modules or that individual hyperedges have been
selected with calibrated error control. Under this constraint the
hyperedge weights no longer have independent closed-form updates; the
factors are fitted by the block-coordinate generalised EM step described
in Section~\ref{sec:em}.

\section{Latent-branching EM estimation}
\label{sec:em}

\subsection{Branching representation}

The EM algorithm treats the source of each observed event as missing
data. For an event $i$ on node $n_i$ at time $t_i$, define
\[
\lambda_i=\lambda_{n_i}(t_i\mid \Hcal_{t_i}).
\]
The event may be a background event, a pairwise offspring of an earlier
event, or a hyperedge-triggered event associated with an active
candidate hyperedge. This is the usual branching idea for self-exciting
point processes, extended here by adding a hyperedge source class.

The background responsibility is
\begin{equation}
p_i^{\rm bg}
=
\frac{\mu_{n_i}}{\lambda_i}.
\label{eq:resp-bg}
\end{equation}
For a previous event $j<i$, the pairwise responsibility is
\begin{equation}
p_{ij}^{\rm pw}
=
\frac{
\alpha_{n_j\to n_i}\phi(t_i-t_j)
}{
\lambda_i
}.
\label{eq:resp-pw}
\end{equation}
For a candidate hyperedge $e$ containing the target node $n_i$, the
hyperedge responsibility is
\begin{equation}
p_{ie}^{\rm he}
=
\frac{
\one\{C_e(t_i)\neq\varnothing\}\alpha_e
\phi\{t_i-t^\star_e(t_i)\}
}{
\lambda_i
}.
\label{eq:resp-he}
\end{equation}
For each event, these responsibilities sum to one after summing
\eqref{eq:resp-pw} over all previous events and \eqref{eq:resp-he} over
all candidate hyperedges that contain the target node. The pairwise and
hyperedge terms therefore compete for the same event-level likelihood
contribution. This competition is central to the identifiability
diagnostics in the simulation study.

\subsection{Expected complete-data objective}

Let
\[
A_e=\sum_{i:n_i\in e} p_{ie}^{\rm he}
\]
be the expected number of events attributed to hyperedge $e$. Let
\[
B_n=\sum_{i:n_i=n}p_i^{\rm bg}
\]
be the expected number of background events on node $n$, and let
\[
P_{a\to n}
=
\sum_{i:n_i=n}\sum_{\substack{j<i\\ n_j=a}}p_{ij}^{\rm pw}
\]
be the expected number of events on node $n$ attributed to previous
events on source node $a$. Up to terms not depending on the parameters,
the expected complete-data objective separates into background,
pairwise, and hyperedge parts:
\begin{align}
Q(\theta)
&=
\sum_{n=1}^N
\left\{
B_n\log\mu_n-\mu_n T
\right\}
\nonumber\\
&\quad+
\sum_{a=1}^N\sum_{n=1}^N
\left\{
P_{a\to n}\log\alpha_{a\to n}
-
\alpha_{a\to n}
\sum_{j:n_j=a}
\frac{1-\exp\{-\beta(T-t_j)\}}{\beta}
\right\}
\nonumber\\
&\quad+
\sum_{e\in\E_{\rm cand}}
\left\{
A_e\log\alpha_e-\alpha_e |e|\Lambda_e
\right\}.
\label{eq:q-function}
\end{align}
An optional $\ell_1$ penalty on independent candidate hyperedge weights
adds $-\eta\sum_{e\in\E_{\rm cand}}\alpha_e$ to
\eqref{eq:q-function}, with $\eta\geq 0$. This penalty is used as a
regularising device; it is not a calibrated post-selection inference
procedure.

\subsection{Closed-form updates for independent weights}

Maximising \eqref{eq:q-function} gives the baseline update
\begin{equation}
\widehat{\mu}_n
=
\frac{B_n}{T}
=
\frac{\sum_{i:n_i=n}p_i^{\rm bg}}{T}.
\label{eq:update-mu}
\end{equation}
For the ordinary pairwise Hawkes weight from source node $a$ to target
node $n$,
\begin{equation}
\widehat{\alpha}_{a\to n}
=
\frac{
P_{a\to n}
}{
\displaystyle
\sum_{j:n_j=a}
\frac{1-\exp\{-\beta(T-t_j)\}}{\beta}
}.
\label{eq:update-pairwise}
\end{equation}
For an independently parameterised candidate hyperedge $e$, the
penalised update is
\begin{equation}
\widehat{\alpha}_e
=
\frac{A_e}{|e|\Lambda_e+\eta}.
\label{eq:update-hyperedge}
\end{equation}
When $\eta=0$, this is the unpenalised maximum-likelihood EM update.
The factor $|e|$ appears because the same hyperedge term contributes to
the intensity of each member node. The exposure $\Lambda_e$ is the
piecewise exposure in \eqref{eq:hyperedge-exposure}, not the naive
exposure obtained by integrating each completion to $T$.

\subsection{CP-constrained generalised EM}

Under the non-negative CP parameterisation
\eqref{eq:cp-param}, the hyperedge weights $\alpha_e$ are coupled
through the factors $F_{vr}$. The independent update
\eqref{eq:update-hyperedge} is then no longer available for each
hyperedge separately. The relevant part of the objective is
\begin{equation}
Q_{\rm he}(F)
=
\sum_{e\in\E_{\rm cand}}
\left\{
A_e\log
\left(
\sum_{r=1}^{R}\prod_{v\in e}F_{vr}
\right)
-
L_e
\left(
\sum_{r=1}^{R}\prod_{v\in e}F_{vr}
\right)
\right\},
\label{eq:q-cp}
\end{equation}
where $L_e=|e|\Lambda_e+\eta$. Holding all other factor entries fixed,
the weight $\alpha_e$ is affine in a single factor entry $F_{vr}$. The
corresponding one-dimensional block subproblem is concave on the
non-negative half-line whenever the candidate hyperedges involving that
entry have non-negative expected counts and exposures. Updating one
factor entry by an exact block maximisation therefore does not decrease
the expected complete-data objective.

The CP-constrained estimator is consequently a generalised EM algorithm:
the E-step is the same as above, while the M-step for the CP factors is
implemented by block-coordinate ascent on \eqref{eq:q-cp}. This
distinction is important. The independent hyperedge weights have the
closed-form update \eqref{eq:update-hyperedge}; the CP factors are
estimated by constrained numerical maximisation of the same EM
objective.

\subsection{Algorithm}

Algorithm~\ref{alg:hth-em} summarises the estimation procedure. The
algorithm is written for independent candidate hyperedge weights. The
CP-constrained version replaces the independent hyperedge update in
Line~\ref{line:he-update} by the block-coordinate step described above.

\begin{algorithm}[!htbp]
\caption{Latent-branching EM for the hyperedge-triggered Hawkes model}
\label{alg:hth-em}
\begin{algorithmic}[1]
\Require Event sequence $\{(t_i,n_i)\}_{i=1}^m$, observation horizon
$T$, candidate set $\E_{\rm cand}$, trigger window $\Delta$, decay rate
$\beta$, penalty $\eta\geq0$.
\State Compute completion sets $C_e$ and piecewise exposures
$\Lambda_e$ for all $e\in\E_{\rm cand}$.
\State Initialise $\mu_n>0$, $\alpha_{a\to n}\geq0$, and
$\alpha_e\geq0$.
\Repeat
  \State Evaluate $\lambda_{n_i}(t_i\mid\Hcal_{t_i})$ for all observed
  events.
  \State Compute background, pairwise, and hyperedge responsibilities
  using \eqref{eq:resp-bg}--\eqref{eq:resp-he}.
  \State Update baseline rates using \eqref{eq:update-mu}.
  \State Update pairwise weights using \eqref{eq:update-pairwise}.
  \State Update independent hyperedge weights using
  \eqref{eq:update-hyperedge}. \label{line:he-update}
  \State Evaluate the observed-data log-likelihood
  \eqref{eq:observed-loglik}.
\Until{the relative increase in log-likelihood is below the convergence
tolerance or the maximum number of iterations is reached.}
\end{algorithmic}
\end{algorithm}

\subsection{Computational remarks}

For the prototype implementation used in the experiments, the dominant
cost is the event-history comparison in the E-step. Without specialised
data structures, evaluating all pairwise event contributions is
quadratic in the number of events. Hyperedge responsibilities add a cost
that depends on the number of active candidate hyperedges and the cost
of computing their anchors. In the reported implementation, completion
sets and exposures are precomputed for the fixed candidate set and
trigger window, while the event-level responsibilities are recomputed at
each EM iteration.

This computational profile is adequate for the moderate event counts
used in the synthetic and exploratory neural analyses, but it is not a
large-scale streaming implementation. Sparse history windows, vectorised
kernel evaluation, and parallel updates would be needed for much larger
records.

\section{Simulation design}
\label{sec:simulation-design}

The synthetic experiments are designed to evaluate estimation behaviour
when the data-generating mechanism is known. This is essential in the
present setting because an observed group-level burst may be explained
either by a genuine hyperedge-triggered component or by ordinary
pairwise excitation. The simulations therefore focus on recovery,
separation, sensitivity, and computation rather than on formal
hypothesis testing.

All synthetic event streams are generated by Ogata thinning
\citep{Ogata1981}. The fitted point-process models are checked with
time-rescaling diagnostics in the simulator validation stage
\citep{Brown2002}. The decay rate $\beta$ is fixed within each
experiment, and the trigger window $\Delta$ is either fixed at its
data-generating value or varied deliberately to examine sensitivity.
The candidate hyperedge set is supplied to the estimator as part of the
model specification. In experiments involving decoy candidates, the
decoys are used to study regularisation and attribution, not to claim
calibrated support recovery.

The synthetic study is organised into four groups. The first group
(Syn01--Syn03) evaluates recovery, regularisation, and EM convergence.
The second group (Syn04--Syn06) studies signal strength, separation from
pairwise Hawkes baselines, and trigger-window sensitivity. The third
group (Syn08--Syn09) examines kernel-timescale effects and practical
identifiability. The final group (Syn07 and Syn10) studies computational
scaling and compares the HTH mechanism with a parameter-matched
third-order surrogate. The main numerical evidence retained in the main
text is reported in Tables~\ref{tab:syn01-recovery}--\ref{tab:syn06-window}
and Figures~\ref{fig:syn01-recovery}--\ref{fig:scaling-baseline}.

For a scalar parameter $\theta$ estimated over $R$ replicated
simulations, the reported bias and root mean squared error are
\begin{equation}
\operatorname{bias}(\widehat{\theta})
=
\frac{1}{R}\sum_{r=1}^{R}\widehat{\theta}_r-\theta_0,
\qquad
\operatorname{RMSE}(\widehat{\theta})
=
\left\{
\frac{1}{R}\sum_{r=1}^{R}
(\widehat{\theta}_r-\theta_0)^2
\right\}^{1/2},
\label{eq:bias-rmse}
\end{equation}
where $\theta_0$ is the data-generating value. The standard deviation is
the empirical standard deviation of $\widehat{\theta}_1,\ldots,
\widehat{\theta}_R$.

For model comparisons, let $\ell_{\HTH}$ and $\ell_{\PW}$ denote the
observed-data log-likelihoods of the hyperedge-triggered and pairwise
models evaluated on the same data. The likelihood gain is
\begin{equation}
\Delta\ell=\ell_{\HTH}-\ell_{\PW}.
\label{eq:delta-loglik}
\end{equation}
When a candidate-count penalty is reported, the statistic is used as a
descriptive BIC-style comparison,
\begin{equation}
\Delta S_{\rm cand}
=
2(\ell_{\HTH}-\ell_{\PW})
-
|\E_{\rm cand}|\log m,
\label{eq:candidate-score-sim}
\end{equation}
where $m$ is the number of events in the evaluation sample. Positive
values indicate that the likelihood gain exceeds this simple
candidate-count penalty. This statistic is not treated as a calibrated
test of individual hyperedges.

The main comparison model is the ordinary pairwise Hawkes process
obtained by removing the hyperedge term from
\eqref{eq:hth-intensity}. In Syn10, the HTH model is also compared with
a parameter-matched third-order surrogate. The purpose of this comparison
is to check whether the pattern-completion mechanism behaves differently
from adding a generic higher-order covariate with a comparable number of
parameters.

\FloatBarrier

\section{Simulation results}
\label{sec:simulation-results}

\subsection{Recovery, regularisation, and convergence}

The first experiment evaluates whether the EM estimator recovers the
parameters in a small setting where the data-generating mechanism is
known. The model has three nodes, observation horizon $T=500$, decay
rate $\beta=1$, trigger window $\Delta=0.5$, and one true candidate
hyperedge $(0,1)$ with weight $\alpha_{(0,1)}=0.4$. Across 25 replicated
simulations, the average number of events was 602.7, with a range from
559 to 697.

\begin{figure}[tbp]
\centering
\safefigure{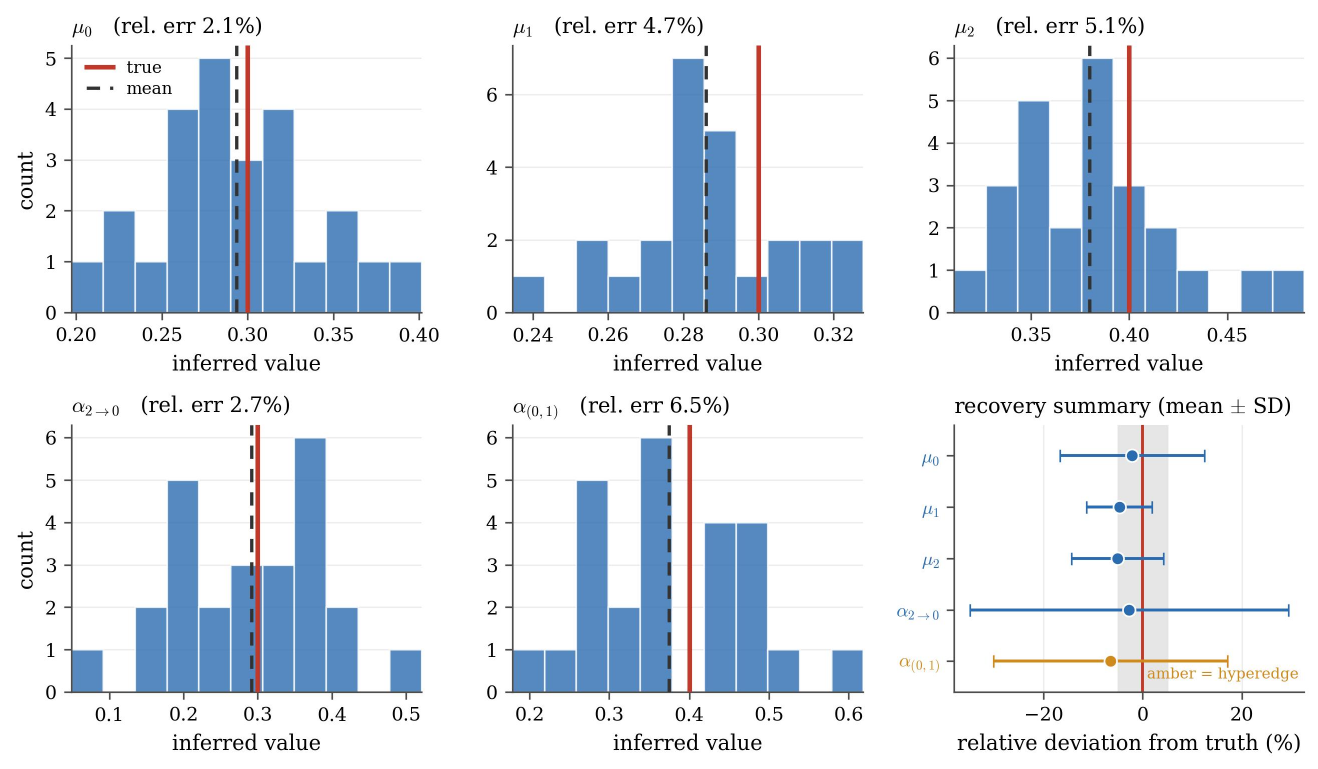}{width=0.86\textwidth}
\caption{Parameter recovery in the replicated synthetic recovery
experiment. The numerical bias and RMSE are reported in
Table~\ref{tab:syn01-recovery}; the figure shows the sampling spread of
the fitted background, pairwise, and hyperedge parameters across
replicated point-process simulations.}
\label{fig:syn01-recovery}
\end{figure}

Table~\ref{tab:syn01-recovery} and
Figure~\ref{fig:syn01-recovery} report the recovery summary. Background
rates are estimated with small negative bias. The selected pairwise
weight $\alpha_{2\to0}$ and the hyperedge weight are also recovered
without large systematic error, although the hyperedge estimate shows a
modest finite-sample downward bias and visible sampling variability.

\begin{table}[tbp]
\centering
\caption{Parameter recovery in Syn01. Bias, standard deviation, and
RMSE are computed over 25 replicated point-process simulations.}
\label{tab:syn01-recovery}
\begin{tabular}{lrrrrr}
\toprule
Parameter & Truth & Mean & Bias & SD & RMSE \\
\midrule
$\mu_0$ & 0.300 & 0.294 & -0.006 & 0.045 & 0.044 \\
$\mu_1$ & 0.300 & 0.286 & -0.014 & 0.020 & 0.024 \\
$\mu_2$ & 0.400 & 0.380 & -0.020 & 0.038 & 0.042 \\
$\alpha_{2\to0}$ & 0.300 & 0.292 & -0.008 & 0.098 & 0.097 \\
$\alpha_{(0,1)}$ & 0.400 & 0.374 & -0.026 & 0.096 & 0.098 \\
\bottomrule
\end{tabular}
\end{table}

The regularisation experiment uses one true candidate and five decoy
candidates. Along the $\ell_1$ path, the fitted true hyperedge weight
decreases gradually but remains active at the largest penalty considered,
whereas the decoy weights are shrunk to numerical zero. In this
benchmark setting, both the AIC and BIC-style criteria select the
largest penalty value, leaving a single active hyperedge. This result is
useful as a sparsity diagnostic, but it should not be read as a general
support-recovery guarantee: the candidate set is fixed in the experiment,
and the penalty path is not calibrated for post-selection inference.

The convergence experiment fits the same dataset from 20 random
initialisations. Each run is allowed 80 EM iterations. The final
log-likelihoods range from $-1125.1207$ to $-1125.1154$, with a spread
of only $0.0053$. This indicates that, for this benchmark dataset, the
algorithm reaches the same likelihood basin from a range of starting
values. The experiment does not rule out local optima in larger or more
ill-conditioned settings, but it shows that the EM updates are stable in
the basic recovery regime.

\subsection{Likelihood separation under signal and null settings}

\begin{figure}[tbp]
\centering
\safefigure{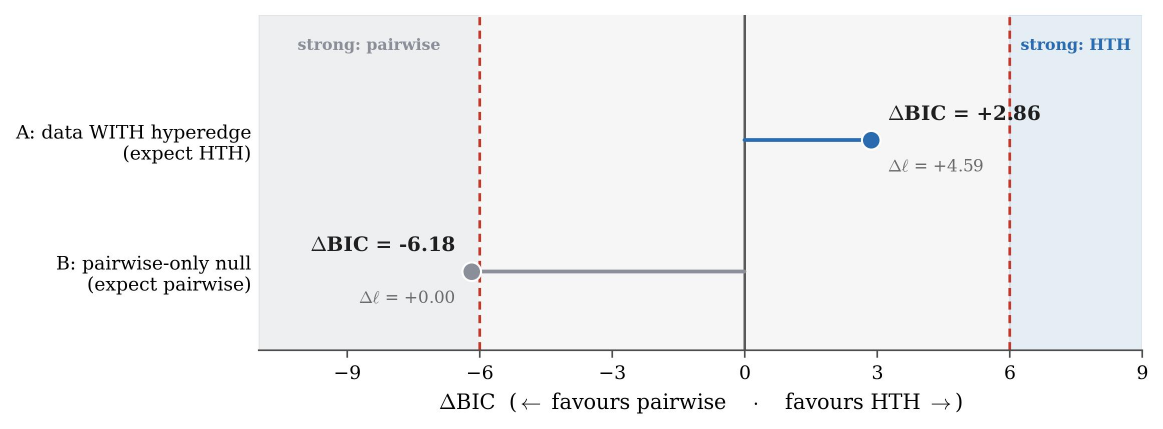}{width=0.72\textwidth}
\caption{Likelihood separation between the pairwise Hawkes model and the
HTH model. The HTH model gains likelihood when the data are generated
with a hyperedge component, while the gain is negligible in the
pairwise-only benchmark null. The historical label $\Delta$BIC inside
the figure denotes the BIC-style score used for this descriptive
comparison; it is not used as a calibrated test of individual
hyperedges.}
\label{fig:syn05-separation}
\end{figure}

The likelihood-separation experiment compares the ordinary pairwise
Hawkes model with the HTH model in two data-generating scenarios. In the
first scenario, the data contain a true hyperedge component. In the
second, the data are generated from a pairwise-only model. Table~\ref{tab:syn05-separation} and
Figure~\ref{fig:syn05-separation} report the likelihood comparison
between the true-hyperedge and pairwise-only benchmark settings.

\begin{table}[tbp]
\centering
\caption{Likelihood separation in Syn05. The BIC-style difference is
used as a descriptive model-comparison statistic, not as a calibrated
test of individual hyperedges.}
\label{tab:syn05-separation}
\begin{tabular}{lrrrr}
\toprule
Scenario & Events & $\Delta\ell$ & BIC-style difference
& Fitted hyperedge weight \\
\midrule
True hyperedge & 552 & 4.587 & 2.861 & 0.318 \\
Pairwise-only null & 485 & 0.00018 & -6.184 & $5.7\times10^{-12}$ \\
\bottomrule
\end{tabular}
\end{table}

When the data contain a hyperedge component, the HTH model gains
log-likelihood and retains a positive fitted hyperedge weight. Under the
pairwise-only null, the likelihood gain is essentially zero and the
fitted hyperedge weight is numerically negligible. This supports the
basic separation behaviour of the estimator in the benchmark scenarios.
It does not establish a general null distribution for the hyperedge
component, which would require a separate testing theory.

\subsection{Trigger-window and kernel-timescale sensitivity}

\begin{figure}[tbp]
\centering
\safefigure{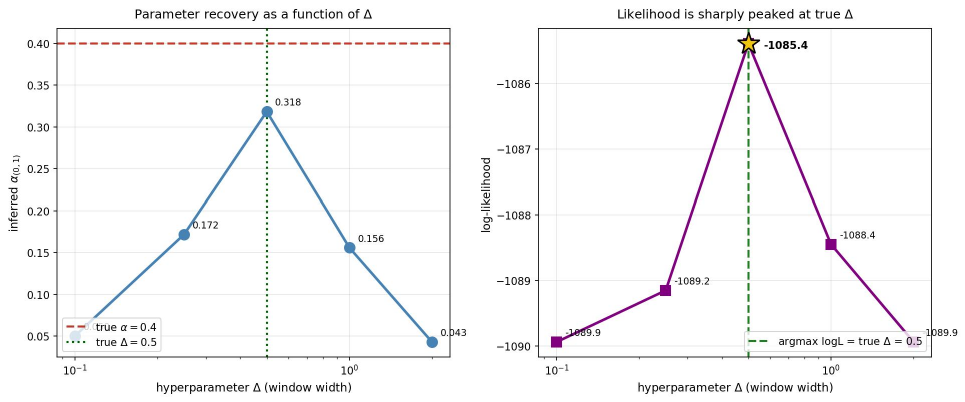}{width=0.72\textwidth}
\caption{Trigger-window sensitivity. The profile log-likelihood is
maximised at the data-generating window $\Delta=0.5$ in this benchmark
experiment, indicating that the event timing contains information about
the pattern-completion scale.}
\label{fig:syn06-window}
\end{figure}

The trigger-window experiment varies the fitted window $\Delta$ while
holding the data fixed. The data-generating value is $\Delta=0.5$. The
profile log-likelihood is maximised at this value, as shown in Table~\ref{tab:syn06-window} and
Figure~\ref{fig:syn06-window}. The fitted hyperedge weight is also
largest at the same window. This suggests that, in the benchmark
setting, the timing information in the event sequence is informative
about the pattern-completion scale.

\begin{table}[tbp]
\centering
\caption{Trigger-window sensitivity in Syn06. The true window is
$\Delta=0.5$.}
\label{tab:syn06-window}
\begin{tabular}{rrr}
\toprule
Fitted window $\Delta$ & Fitted hyperedge weight & Log-likelihood \\
\midrule
0.10 & 0.050 & -1089.941 \\
0.25 & 0.172 & -1089.153 \\
0.50 & 0.318 & -1085.388 \\
1.00 & 0.156 & -1088.449 \\
2.00 & 0.043 & -1089.941 \\
\bottomrule
\end{tabular}
\end{table}

The kernel-timescale ablation varies the decay rate $\beta$ while
keeping the true hyperedge weight fixed at $0.4$. Figure~\ref{fig:identifiability}(a)--(b) reports the fitted hyperedge weights
over 25 seeds for each value of $\beta$. The main pattern is not a monotone bias but a
loss of precision at faster timescales. As $\beta$ increases, the
standard deviation and RMSE of the hyperedge estimate increase
substantially. This is consistent with the identifiability interpretation
of the model: when excitation decays too quickly, fewer events carry
clear temporal evidence for separating pairwise and hyperedge
responsibilities. The detailed numerical values are shown directly in
Figure~\ref{fig:identifiability} rather than repeated in a separate
main-text table.

\subsection{Identifiability and candidate effects}

\begin{figure}[!htbp]
\centering
\safefigure{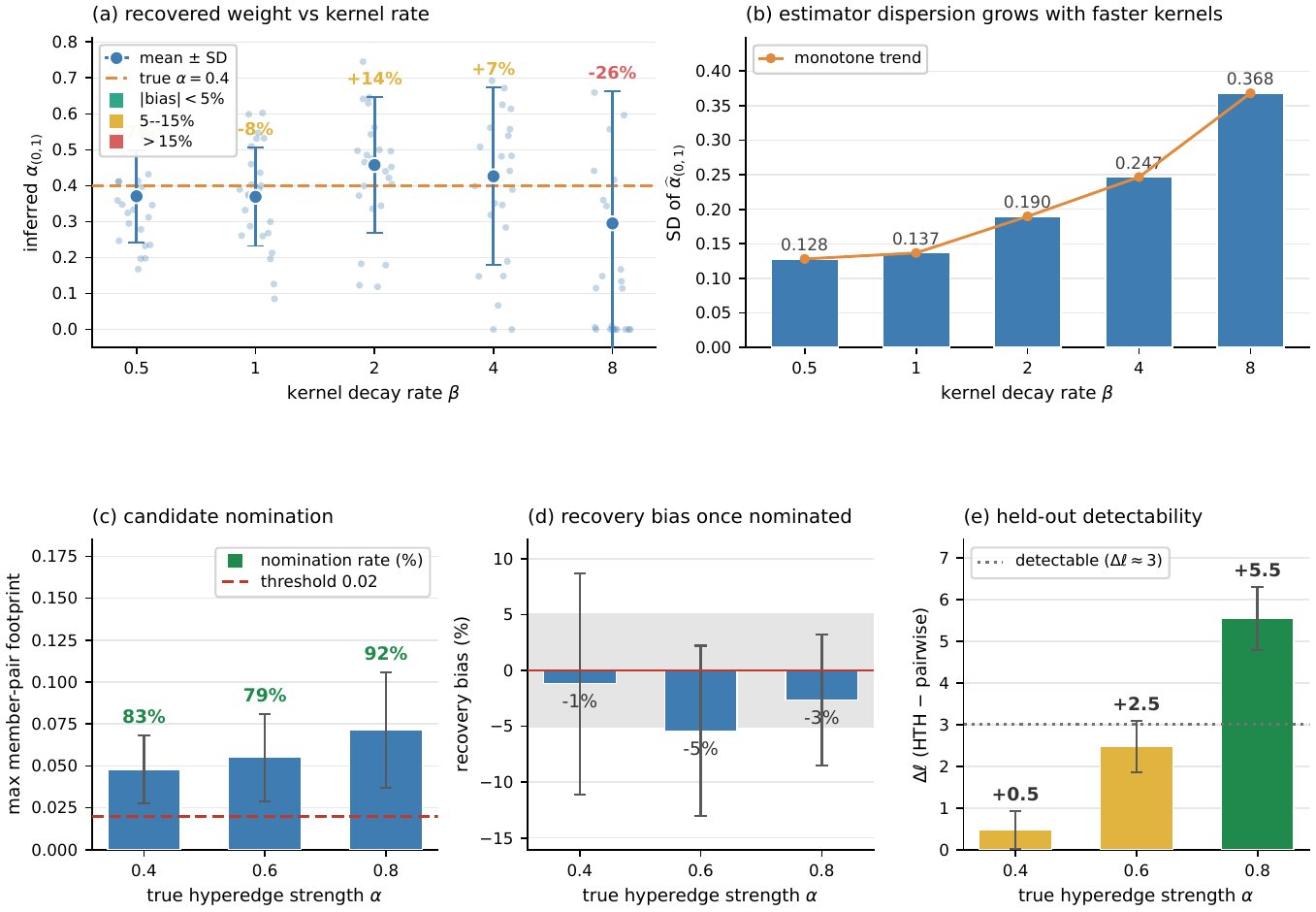}{width=0.98\textwidth}
\caption{Identifiability diagnostics. Panels (a)--(b) summarise the
kernel-timescale ablation, showing how recovery and estimator dispersion
change with the decay rate. Panels (c)--(e) summarise the candidate and
likelihood diagnostics, showing that nomination alone is not sufficient
for stable separation from pairwise structure.}
\label{fig:identifiability}
\end{figure}

The identification diagnostic, summarised in
Figure~\ref{fig:identifiability}(c)--(e), examines a three-node true hyperedge
$(0,1,2)$ over three signal strengths. The candidate nomination rate is
high even at the weakest signal, but the likelihood gain remains small.
This shows that candidate nomination is not the only bottleneck:
a nominated candidate may still be hard to distinguish from pairwise
structure if the higher-order contribution is weak.

The result is useful precisely because it is not uniformly positive.
It indicates that the model can recover the hyperedge weight reasonably
well in some settings while still failing to obtain a decisive
likelihood separation at weaker signal levels. This is the practical
form of the pairwise--hyperedge confounding problem: the same local
co-firing events can be partially explained by ordinary pairwise
excitation and by the group-completion term.

\subsection{Computational scaling and interaction baseline}

\begin{figure}[!htbp]
\centering
\safefigure{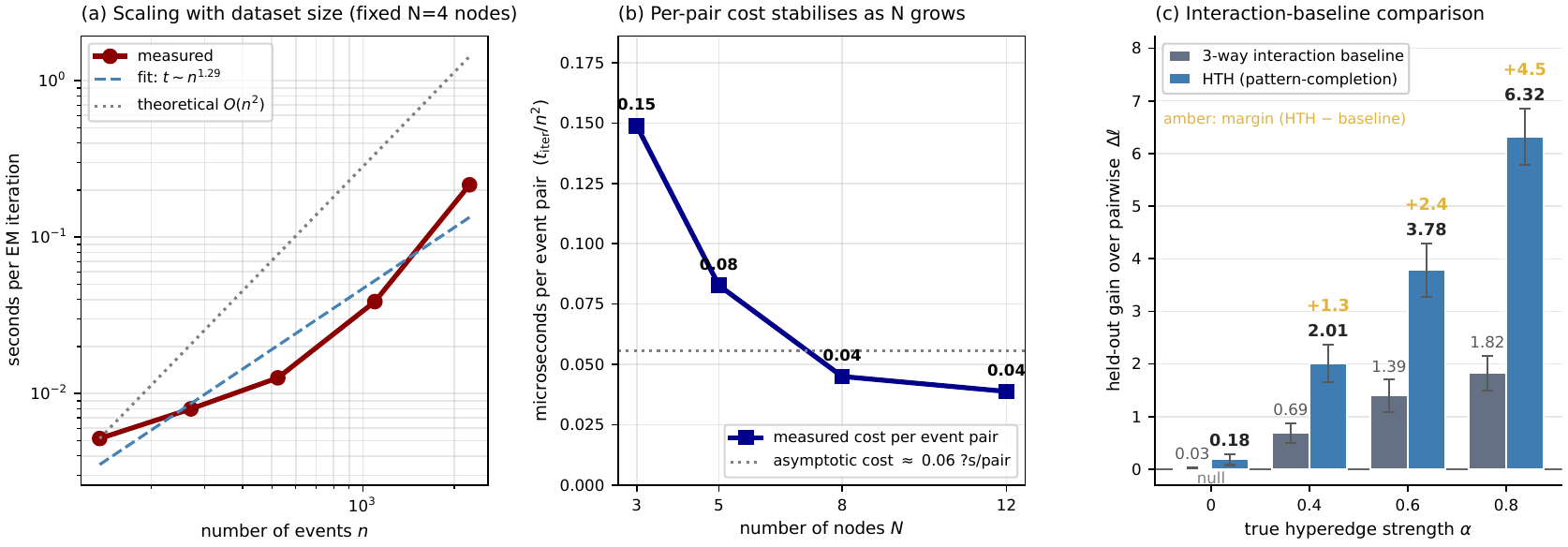}{width=0.98\textwidth}
\caption{Computation and interaction-baseline comparison. Panels
(a)--(b) show event-count scaling and per-pair computational cost for
the prototype implementation. Panel (c) compares the HTH mechanism with
a parameter-matched third-order surrogate; the HTH margin increases with
the true pattern-completion strength.}
\label{fig:scaling-baseline}
\end{figure}

The prototype implementation is dominated by event-history comparisons
in the E-step. The event-count scaling experiment, summarised in
Figure~\ref{fig:scaling-baseline}(a)--(b), varies the observation horizon and
records the time for one representative EM step. A quadratic
fit to runtime as a function of the number of events gives
$R^2=0.999$, compared with $R^2=0.915$ for a linear fit. On the three
largest event counts, the log--log slope is approximately $1.95$. These
numbers are consistent with the expected near-quadratic event-count
cost of the current implementation.

Finally, Syn10 compares the HTH model with a parameter-matched
third-order surrogate. The surrogate is included to check whether the
gain comes merely from adding another higher-order covariate, rather
than from the pattern-completion mechanism itself. Figure~\ref{fig:scaling-baseline}(c) reports the mean likelihood gains over
24 seeds.

Both models have small positive gains in the null setting, which is
expected for nested or flexible comparisons on finite samples. The
important pattern in Figure~\ref{fig:scaling-baseline}(c) is that the
HTH margin grows with the true pattern-completion strength. This supports the use of the
most-recent-anchor mechanism as a specific temporal model rather than as
an arbitrary third-order adjustment.

\FloatBarrier

\section{Exploratory neural applications}
\label{sec:real-data}

\subsection{Datasets and purpose}

The real-data analyses use three publicly available neural spike-train
datasets from the CRCNS data-sharing platform: the ret-1 retinal
ganglion-cell dataset, the pvc-3 primary visual-cortex dataset, and the
pvc-11 macaque primary visual-cortex dataset. These recordings are used
to illustrate how the fitted HTH model behaves on event streams with
different candidate-set sizes and different levels of apparent
higher-order temporal structure.

The purpose of this section is predictive and exploratory. The
simulation experiments provide the controlled setting in which the true
higher-order mechanism is known. In real neural recordings, the
generating mechanism is not observed, candidate sets are constructed
from the data, and external covariates may also affect co-firing
patterns. The analysis below should therefore be read as a held-out
model-comparison diagnostic. A positive score indicates improved
predictive fit under the fitted model class; it does not constitute a
calibrated test or a biological validation of an individual hyperedge.

\subsection{Candidate-penalised held-out score}

For each dataset and candidate-set size, an ordinary pairwise Hawkes
model and an HTH model are fitted and compared on held-out evaluation
windows. Let $\ell_{\HTH}^{\rm test}$ and $\ell_{\PW}^{\rm test}$
denote the held-out log-likelihoods of the two models on the same test
window, and let $m_{\rm test}$ be the number of test events. We report
the candidate-penalised held-out score
\begin{equation}
\Delta S_{\rm cand}^{\rm test}
=
2\left(
\ell_{\HTH}^{\rm test}
-
\ell_{\PW}^{\rm test}
\right)
-
|\E_{\rm cand}|\log m_{\rm test}.
\label{eq:realdata-score}
\end{equation}
This score is a descriptive comparison between fitted predictive models.
Positive values mean that the held-out likelihood gain exceeds the
simple candidate-count penalty. The score should not be interpreted as a
formal BIC theorem for the selected candidates or as a hypothesis test
for individual hyperedges.

\subsection{Results across candidate-set sizes}

\begin{figure}[tbp]
\centering
\safefigure{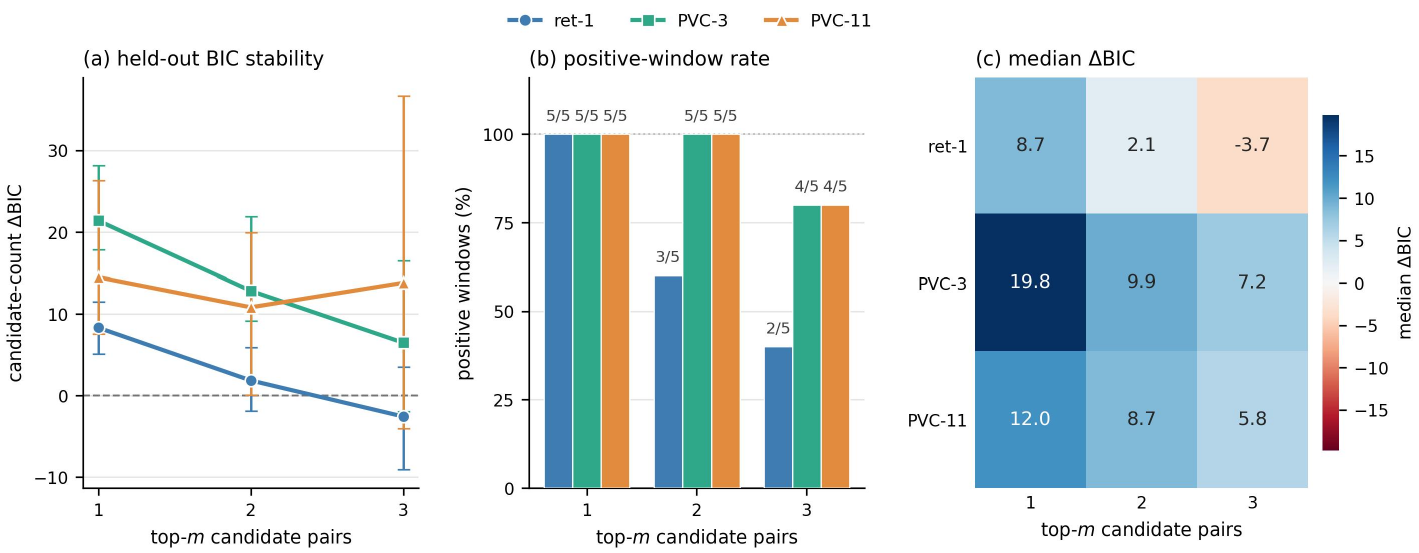}{width=0.92\textwidth}
\caption{Exploratory neural-data summary. The panels show the
candidate-penalised held-out score across candidate-set sizes and
datasets. The historical $\Delta$BIC labels in the figure refer to the
score defined in \eqref{eq:realdata-score}. Positive values indicate
improved held-out predictive fit after the candidate-count penalty, but
the score is not a calibrated test of individual biological hyperedges.}
\label{fig:realdata-summary}
\end{figure}

Table~\ref{tab:realdata-summary} and
Figure~\ref{fig:realdata-summary} summarise the held-out comparisons
over five evaluation windows for each dataset and candidate-set size.
The ret-1 dataset shows the strongest dependence on candidate sparsity:
the score is positive in all five windows for the sparsest candidate set,
but becomes mixed and then negative as more candidates are admitted. The
two visual-cortex datasets show more persistent positive scores across
the reported candidate-set sizes, although the scores still vary across
windows and candidate choices.

\begin{table}[tbp]
\centering
\caption{Exploratory neural-data summary. Positive windows count how
often the candidate-penalised held-out score in
\eqref{eq:realdata-score} is positive across five evaluation windows.}
\label{tab:realdata-summary}
\begin{tabular}{llrrr}
\toprule
Dataset & Candidate size & Positive windows
& Mean $\Delta S_{\rm cand}^{\rm test}$
& Median $\Delta S_{\rm cand}^{\rm test}$ \\
\midrule
ret-1 & top-$1$ & 5/5 & 8.322 & 8.677 \\
ret-1 & top-$2$ & 3/5 & 1.856 & 2.132 \\
ret-1 & top-$3$ & 2/5 & -2.565 & -3.692 \\
PVC-3 area17 & top-$1$ & 5/5 & 21.391 & 19.801 \\
PVC-3 area17 & top-$2$ & 5/5 & 12.822 & 9.887 \\
PVC-3 area17 & top-$3$ & 4/5 & 6.486 & 7.232 \\
PVC-11 monkey2 & top-$1$ & 5/5 & 14.479 & 12.020 \\
PVC-11 monkey2 & top-$2$ & 5/5 & 10.821 & 8.732 \\
PVC-11 monkey2 & top-$3$ & 4/5 & 13.814 & 5.840 \\
\bottomrule
\end{tabular}
\end{table}

The pattern is consistent with the simulation study. The HTH component
can improve predictive fit when the candidate set captures useful
short-window group structure. At the same time, interpretation becomes
less stable as the candidate set expands, because several candidates may
compete to explain the same co-firing evidence. The cortical recordings
show stronger candidate-penalised held-out gains than the retinal
recording in this analysis, but this should not be read as evidence that
individual cortical hyperedges have been confirmed. The appropriate
conclusion is narrower: in these data and under the fitted candidate
sets, the HTH component sometimes provides predictive information beyond
the pairwise Hawkes baseline, while the magnitude and sign of the gain
remain sensitive to candidate construction.

\section{Discussion}
\label{sec:discussion}

\subsection{Interpretation of fitted HTH components}

The HTH model decomposes the conditional intensity into background
activity, ordinary pairwise Hawkes excitation, and a
hyperedge-triggered source class. The EM algorithm estimates how much
event-level likelihood is assigned to these components. This
decomposition is useful for modelling and diagnosis, but it is not a
direct observation of mechanism. A fitted hyperedge weight should be
interpreted as a model component that improves the likelihood under a
specified candidate set, trigger window, and kernel timescale.

The most-recent-anchor convention is central to this interpretation. A
completion of a candidate hyperedge creates one active anchor, and a
later completion of the same candidate group replaces it. The
higher-order component is therefore different from a linear Hawkes term
that accumulates all past events. The piecewise compensator follows from
this definition. Using the ordinary accumulated Hawkes exposure would
estimate a different model and would overcount the contribution of
superseded anchors.

\subsection{Pairwise--hyperedge confounding}

The main statistical challenge is attribution. A local burst of events
may increase the HTH likelihood, but the same burst may also be
partially explained by ordinary pairwise excitation. This difficulty is
expected in neural population data, where weak pairwise correlations can
coexist with strongly structured population states
\citep{Schneidman2006}. It also appears in the controlled simulations:
a true candidate can be nominated frequently and estimated with moderate
bias, while the likelihood gain remains too small for stable separation
from the pairwise baseline at weaker signal strengths.

For this reason, the simulation experiments are the primary evidence for
estimation behaviour. They show when recovery and separation are
possible because the data-generating mechanism is known. The neural
applications answer a different question: whether the fitted
higher-order component improves held-out prediction relative to a
pairwise Hawkes baseline. They do not establish a null distribution for
selected hyperedges, and they should not be read as confirmatory
evidence for biological higher-order circuits.

\subsection{Candidate sets, regularisation, and model comparison}

The candidate set is part of the fitted model. It restricts the number
of possible higher-order components and makes likelihood evaluation
computationally feasible. Sparse regularisation and low-dimensional
structure are common tools in high-dimensional Hawkes modelling
\citep{Hansen2015,Bacry2020SparseLowRank}, and they play a similar
stabilising role here. However, the candidate generator, the
regularisation path, and the candidate-penalised held-out score do not
provide calibrated post-selection inference.

This distinction matters for reporting real-data results. A positive
candidate-penalised held-out score means that, for the chosen candidate
set, the HTH model predicts the held-out events better than the
pairwise Hawkes baseline after a simple complexity penalty. It does not
mean that any particular candidate has passed a formal support test.
Formal testing is already delicate for non-negative Hawkes interactions
because zero interaction parameters lie on the boundary of the parameter
space \citep{Lotz2024}. The HTH setting adds further complications:
candidate sets may be data-dependent, and the higher-order component is
a pattern-completion source rather than an ordinary pairwise offspring
process.

\subsection{Practical guidance}

The experiments suggest four practical guidelines. First, the candidate
set should be kept sparse and scientifically interpretable. Adding many
weakly exposed candidates can increase flexibility without improving
identifiability. Second, fitted hyperedge weights should be interpreted
together with exposure, held-out likelihood gain, and sensitivity to the
candidate-set size. A positive weight alone is not sufficient evidence
for a stable higher-order effect. Third, trigger-window choice should be
checked empirically. In the benchmark simulation, the likelihood profile
is maximised near the data-generating window, but this behaviour should
not be assumed in applications. Fourth, kernel timescale affects
precision. Faster decay can reduce the amount of temporal evidence
available for separating pairwise and hyperedge responsibilities.

These guidelines are deliberately conservative. The HTH model is most
useful when the scientific question concerns temporally organised group
completion and when simulation or held-out diagnostics show that the
higher-order component adds information beyond pairwise excitation. It
is less appropriate as an automatic detector of biological hyperedges
from a single fitted likelihood gain.

\subsection{Limitations and future work}

The first limitation is the absence of a formal calibration procedure
for selected HTH components. Developing such a procedure would require a
separate analysis of boundary effects, data-dependent candidate sets,
and the most-recent-anchor source class. Parametric simulation under
structured pairwise null models, repeated data splitting, or
selective-inference ideas may be useful directions, but they are outside
the estimation method developed here.

The second limitation is computational. The prototype implementation has
near-quadratic event-count scaling over the tested range because it
recomputes event-history contributions in the E-step. This is adequate
for the moderate synthetic and exploratory neural records studied here,
but larger event streams would require sparse history windows,
vectorised kernel evaluation, or parallel computation.

The third limitation is modelling scope. The experiments use a fixed
exponential decay rate within each fitted model. Node-specific or
edge-specific decay rates may be more appropriate for heterogeneous
event streams. The neural applications also do not include stimulus
covariates. Stimulus-driven co-activation can resemble network-driven
group interaction, so applications to stimulus-response data should
include appropriate covariate structure before assigning mechanistic
meaning to fitted hyperedge terms.
\section{Conclusion}
\label{sec:conclusion}

This paper introduced a hyperedge-triggered Hawkes model for
continuous-time event streams. The model augments ordinary pairwise
Hawkes excitation with a short-window pattern-completion component: a
candidate hyperedge becomes active when all of its member nodes fire
within a prescribed temporal window. The most-recent-anchor convention
makes this component non-accumulating, so the likelihood requires a
piecewise compensator for superseded anchors rather than the ordinary
integral over all past completions.

The resulting estimator is a latent-branching EM algorithm. Background,
pairwise, and hyperedge responsibilities are computed at the event
level, and independently parameterised candidate hyperedge weights admit
closed-form M-step updates. When a non-negative CP parameterisation is
used to reduce the number of hyperedge parameters, the same expected
complete-data objective can be fitted by a block-coordinate
generalised EM step.

The simulation study shows that the estimator recovers higher-order
excitation in informative regimes, converges stably in the benchmark
settings, identifies the correct trigger window in a controlled
experiment, and exhibits the expected near-quadratic event-count scaling
of the prototype implementation. It also identifies the main limitation:
pairwise and hyperedge components may compete for the same co-firing
evidence, so candidate nomination and positive fitted weights are not by
themselves sufficient for stable higher-order interpretation.

The neural spike-train analyses illustrate the model's use as a
predictive diagnostic. Candidate-penalised held-out gains can be
positive, especially in sparse candidate regimes, but they are sensitive
to candidate construction and should not be interpreted as calibrated
discoveries of individual biological hyperedges. The proposed method
therefore provides a likelihood-based framework for studying
higher-order temporal dependence, while formal post-selection testing and
biological validation remain separate problems.
\section*{Data availability statement}

The raw neural recordings used in the exploratory applications are
available from the CRCNS data-sharing platform subject to its access
conditions. The ret-1, pvc-3, and pvc-11 datasets are public CRCNS
datasets. Code, reproducibility scripts, synthetic-result files, and
processed real-data summaries are available at
\url{https://github.com/Hanii0210/hypergraph-hawkes}.

\section*{Disclosure statement}

No potential conflict of interest is reported.

% ----------------------------------------------------------------------
% Temporary bibliography for compilation.
% This section will be replaced by the final reference list after the
% manuscript text is complete.
% ----------------------------------------------------------------------


\begin{thebibliography}{99}

\bibitem[Hawkes(1971)]{Hawkes1971}
Hawkes, A. G. (1971).
Spectra of some self-exciting and mutually exciting point processes.
\emph{Biometrika}, 58(1), 83--90.

\bibitem[Ozaki(1979)]{Ozaki1979}
Ozaki, T. (1979).
Maximum likelihood estimation of Hawkes' self-exciting point processes.
\emph{Annals of the Institute of Statistical Mathematics}, 31, 145--155.

\bibitem[Ogata(1978)]{Ogata1978}
Ogata, Y. (1978).
The asymptotic behaviour of maximum likelihood estimators for stationary
point processes.
\emph{Annals of the Institute of Statistical Mathematics}, 30, 243--261.

\bibitem[Ogata(1981)]{Ogata1981}
Ogata, Y. (1981).
On Lewis' simulation method for point processes.
\emph{IEEE Transactions on Information Theory}, 27(1), 23--31.

\bibitem[Brown et~al.(2002)]{Brown2002}
Brown, E. N., Barbieri, R., Ventura, V., Kass, R. E., and Frank, L. M.
(2002).
The time-rescaling theorem and its application to neural spike train
data analysis.
\emph{Neural Computation}, 14(2), 325--346.

\bibitem[Truccolo et~al.(2005)]{Truccolo2005}
Truccolo, W., Eden, U. T., Fellows, M. R., Donoghue, J. P., and
Brown, E. N. (2005).
A point process framework for relating neural spiking activity to
spiking history, neural ensemble, and extrinsic covariate effects.
\emph{Journal of Neurophysiology}, 93(2), 1074--1089.

\bibitem[Schneidman et~al.(2006)]{Schneidman2006}
Schneidman, E., Berry, M. J., Segev, R., and Bialek, W. (2006).
Weak pairwise correlations imply strongly correlated network states in a
neural population.
\emph{Nature}, 440, 1007--1012.

\bibitem[Pillow et~al.(2008)]{Pillow2008}
Pillow, J. W., Shlens, J., Paninski, L., Sher, A., Litke, A. M.,
Chichilnisky, E. J., and Simoncelli, E. P. (2008).
Spatio-temporal correlations and visual signalling in a complete
neuronal population.
\emph{Nature}, 454, 995--999.

\bibitem[Kass et~al.(2011)]{Kass2011}
Kass, R. E., Kelly, R. C., and Loh, W.-L. (2011).
Assessment of synchrony in multiple neural spike trains using loglinear
point process models.
\emph{The Annals of Applied Statistics}, 5(2B), 1262--1292.

\bibitem[Veen and Schoenberg(2008)]{VeenSchoenberg2008}
Veen, A., and Schoenberg, F. P. (2008).
Estimation of space--time branching process models in seismology using
an EM-type algorithm.
\emph{Journal of the American Statistical Association}, 103(482),
614--624.

\bibitem[Lewis and Mohler(2011)]{LewisMohler2011}
Lewis, E., and Mohler, G. (2011).
A nonparametric EM algorithm for multiscale Hawkes processes.
Preprint.

\bibitem[Zhou et~al.(2013)]{Zhou2013}
Zhou, K., Zha, H., and Song, L. (2013).
Learning triggering kernels for multi-dimensional Hawkes processes.
In \emph{Proceedings of the 30th International Conference on Machine
Learning}, 1301--1309.

\bibitem[Hansen et~al.(2015)]{Hansen2015}
Hansen, N. R., Reynaud-Bouret, P., and Rivoirard, V. (2015).
Lasso and probabilistic inequalities for multivariate point processes.
\emph{Bernoulli}, 21(1), 83--143.

\bibitem[Bacry et~al.(2020)]{Bacry2020SparseLowRank}
Bacry, E., Bompaire, M., Ga\"iffas, S., and Poulsen, S. V. (2020).
Sparse and low-rank multivariate Hawkes processes.
\emph{Journal of Machine Learning Research}, 21, 1--32.

\bibitem[Xu et~al.(2016)]{Xu2016}
Xu, H., Farajtabar, M., and Zha, H. (2016).
Learning Granger causality for Hawkes processes.
In \emph{Proceedings of the 33rd International Conference on Machine
Learning}, 1717--1726.

\bibitem[Eichler et~al.(2017)]{Eichler2017}
Eichler, M., Dahlhaus, R., and Dueck, J. (2017).
Graphical modeling for multivariate Hawkes processes with nonparametric
link functions.
\emph{Journal of Time Series Analysis}, 38(2), 225--242.

\bibitem[Embrechts and Kirchner(2018)]{EmbrechtsKirchner2018}
Embrechts, P., and Kirchner, M. (2018).
Hawkes graphs.
\emph{Theory of Probability and Its Applications}, 62(1), 132--156.

\bibitem[Bonnet(2021)]{Bonnet2021}
Bonnet, A. (2021).
Maximum likelihood estimation for Hawkes processes with self-excitation
or inhibition.
Preprint.

\bibitem[Bonnet et~al.(2023)]{Bonnet2023}
Bonnet, A., Martinez Herrera, M., and Sangnier, M. (2023).
Inference of multivariate exponential Hawkes processes with inhibition
and application to neuronal activity.
\emph{Statistics and Computing}, 33, Article 91.

\bibitem[Lotz(2024)]{Lotz2024}
Lotz, J. (2024).
A sparsity test for multivariate Hawkes processes.
Preprint.

\bibitem[Bonnet et~al.(2025)]{BonnetTesting2025}
Bonnet, A., Dion-Blanc, C., and Sadeler Perrin, C. (2025).
Testing procedures based on maximum likelihood estimation for marked
Hawkes processes.
Preprint.

\bibitem[Battiston et~al.(2020)]{Battiston2020}
Battiston, F., Cencetti, G., Iacopini, I., Latora, V., Lucas, M.,
Patania, A., Young, J.-G., and Petri, G. (2020).
Networks beyond pairwise interactions: structure and dynamics.
\emph{Physics Reports}, 874, 1--92.

\bibitem[Giusti et~al.(2016)]{Giusti2016}
Giusti, C., Ghrist, R., and Bassett, D. S. (2016).
Two's company, three (or more) is a simplex: algebraic-topological tools
for understanding higher-order structure in neural data.
\emph{Journal of Computational Neuroscience}, 41, 1--14.

\bibitem[Zhang et~al.(2023)]{Zhang2023HigherOrder}
Zhang, Y., Lucas, M., and Battiston, F. (2023).
Higher-order interactions shape collective dynamics differently in
hypergraphs and simplicial complexes.
\emph{Nature Communications}, 14, Article 1605.

\bibitem[Kolda and Bader(2009)]{KoldaBader2009}
Kolda, T. G., and Bader, B. W. (2009).
Tensor decompositions and applications.
\emph{SIAM Review}, 51(3), 455--500.

\bibitem[Zhang et~al.(2014)]{ZhangRet1}
Zhang, Y. F., Asari, H., and Meister, M. (2014).
Multi-electrode recordings from retinal ganglion cells.
CRCNS.org. doi:10.6080/K0RF5RZT.

\bibitem[Blanche(2009)]{BlanchePvc3}
Blanche, T. (2009).
Multi-neuron recordings in primary visual cortex.
CRCNS.org. doi:10.6080/K0MW2F2J.

\bibitem[Kohn and Smith(2016)]{KohnSmithPvc11}
Kohn, A., and Smith, M. A. (2016).
Utah array extracellular recordings of spontaneous and visually evoked
activity from anesthetized macaque primary visual cortex.
CRCNS.org. doi:10.6080/K0NC5Z4X.

\end{thebibliography}
\end{document}